\def\bq{\begin{equation}}
\def\eq{\end{equation}}
\def\bqa{\begin{eqnarray}}
\def\eqa{\end{eqnarray}}
\def\bqb{\begin{eqnarray*}}
\def\eqb{\end{eqnarray*}}
\documentstyle[12pt]{article}\textwidth 16cm
\def\pr#1#2#3{ Phys. Rev. ${\bf{#1}}$ (#2) #3}

\def\pl#1#2#3{ Phys. Lett. ${\bf{#1}}$ (#2) #3 }
\def\prep#1#2#3{ Phys. Rep. ${\bf{#1}}$ (#2) #3}
\def\np#1#2#3{ Nucl. Phys. ${\bf{#1}}$ (#2) #3}
\def\zp#1#2#3{ Z. Phys. ${\bf{#1}}$ (#2) #3}

\def\etal{{\it et.al.\/}}

\global\nulldelimiterspace = 0pt

\def\Bsl{\hbox{/\kern-.6700em$B$}} 
\def\Dsl{\hbox{/\kern-.6700em$D$}} 
\def\Wsl{\hbox{/\kern-.6700em$W$}} 

\def\roughly#1{\mathrel{\raise.3ex
    \hbox{$#1$\kern-.75em\lower1ex\hbox{$\sim$}}}}
\def\lsim{\roughly<}
\def\gsim{\roughly>}

\def\mh2{m^2_H}

\begin{document}
\pagenumbering{arabic}
\thispagestyle{empty}
\hspace {-0.8cm} PM/95-39\\
\hspace {-0.8cm} CPT-95/P.3247\\
\hspace {-0.8cm} September 1995\\
\vspace {0.8cm}\\

\begin{center}
{\Large\bf Anomalous $Z'$ effects in the WW channel at NLC} \\

 \vspace{1.8cm}
{\large  P. Chiappetta$^a$, F.M. Renard$^b$ and C.
Verzegnassi$^c$}
\vspace {1cm}  \\
$^a$Centre
de Physique Th\'{e}orique,
UPR 7061,\\
CNRS Luminy, Case 907, F-13288 Marseille Cedex 9.\\
\vspace{0.2cm}
$^b$Physique
Math\'{e}matique et Th\'{e}orique,
CNRS-URA 768,\\
Universit\'{e} de Montpellier II,
 F-34095 Montpellier Cedex 5.\\
\vspace{0.2cm}
$^c$ Dipartimento di Fisica,
Universit\`{a} di Lecce \\
CP193 Via Arnesano, I-73100 Lecce, \\
and INFN, Sezione di Lecce, Italy.\\

\vspace{1.5cm}

 {\bf Abstract}
\end{center}
\noindent
We consider the virtual signals of a $Z'$ of very general type in the
process $e^+e^-\to W^+W^-$ at a future linear collider (NLC). We show
that possible deviations from the SM predictions in this channel are
related to similar deviations in the purely leptonic one in a way that
is only characteristic of this $Z'$ model, and not in general of possible
competitor models with anomalous gauge couplings.

\vspace{1cm}

\setcounter{page}{0}
\def\thefootnote{\arabic{footnote}}
\setcounter{footnote}{0}
\clearpage

\section{Introduction}

The existence of one extra $ Z (\equiv Z'$),
with a mass not far from the
conventional electroweak breaking scale of a few hundred GeV, has been
naturally expected in several, theoretically appealing, models whose
motivations vary from those of the "historical" $SO(10)$, $E_6$
proposals \cite {group} to those of a number of more recent publications
\cite {group1}. Owing to the lack of theoretical predictions on the
mass of this extra $Z (\equiv M_{Z'}$), both direct production and
virtual signals have been considered in the literature. In particular,
{}from various analyses on $Z$ resonance \cite{ZZp} (where, strictly
speaking, no information can be derived on $M_{Z'}$, with the
remarkable exception of a very small number of "constrained" models)
the conclusion emerges that the $Z'$ mixing with $Z$ can be ignored for
all theoretical analyses at higher energy $e^+e^-$ colliders where
either final fermionic channels \cite{Zplep2} or $WW$ channel with no
initial lepton polarization \cite{Zpww} are considered. Direct
production at present and future colliders has also been examined in
several papers \cite{Zpdir}. This might certainly be relevant for
hadronic colliders i.e. for Tevatron and LHC. Since the present
experimental limits are already of the order of five hundred GeV
\cite{Zpfnal}, only virtual effects have some interest at the future
$e^+e^-$ colliders, in particular at LEP2 and at a 500 GeV linear
collider (NLC) \cite{nlc}, that we shall consider in this paper as
the realistic possibility of a not too far future. In particular, a
previous realistic theoretical analysis of virtual effects in the
fermionic final channel at NLC that included QED effects has been
performed for the specific case of an $E_6$ generated $Z'$ \cite{Zpqed}.
The aim of this short paper is that of presenting a preliminary
generalization of the analysis of ref.\cite{Zpqed} which treats the
$Z'$ couplings to fermions as free parameters (assuming charged lepton
universality) and also considers the possibility of a final $WW$
state, without initial lepton polarization. As we shall show, the
combination of the $WW$ channel with the purely leptonic ones would
evidentiate special correlations between the different effects that
would be an intrinsic characteristics of any model with such a general
$Z'$.\par
Following our previous remarks on the $Z-Z'$ mixing, the virtual
effects of a $Z'$ in the process of $e^+e^-$ annihilation into leptons
or $W$ pairs at c.m. squared energy $\equiv q^2$
can be described, at tree
level, by adding to the Standard Model
$\gamma, Z$ and $\nu$ exchanges the diagram
with $Z'$ exchange. The overall effect in the relevant scattering
amplitudes will be summarized by the following expressions:

\bq A^{(0)}_{ll}(q^2) =  A^{(0)(\gamma, Z)}_{ll}(q^2)
+ A^{(0)(Z')}_{ll}(q^2)  \eq

\bq A^{(0)}_{lW}(q^2) =  A^{(0)(\gamma, Z, \nu)}_{lW}(q^2)
+ A^{(0)(Z')}_{lW}(q^2)  \eq

\noindent
where $l=e, \mu, \tau$ and we assume universal $Z'll$ couplings. In
eq.(1) one should read:

\bq A^{(0)(\gamma)}_{ll}(q^2) = {ie^2_0\over q^2}\bar
v_l\gamma_{\mu}u_l \bar u_l\gamma^{\mu}v_l \eq

\bq A^{(0)(Z)}_{ll}(q^2) =  {i\over
q^2-M^2_{0Z}}({g^2_0\over4c^2_0})\bar
v_l\gamma_{\mu}(g^{(0)}_{Vl}-\gamma_5g^{(0)}_{Al})u_l
\bar u_l\gamma^{\mu}(g^{(0)}_{Vl}-\gamma_5g^{(0)}_{Al})v_l \eq

\bq A^{(0)(Z')}_{ll}(q^2) =   {i\over
q^2-M^2_{0Z'}}({g^2_0\over4c^2_0})\bar
v_l\gamma_{\mu}(g^{'(0)}_{Vl}-\gamma_5g^{'(0)}_{Al})u_l
\bar u_l\gamma^{\mu}(g^{'(0)}_{Vl}-\gamma_5g^{'(0)}_{Al})v_l\eq

\noindent
($e^2_0=g^2_0s^2_0$, $s^2_0=1-c^2_0$). Note the choice of
normalization in eq.(4): $g^{(0)}_{Al}=-{1\over2}$ and $g^{(0)}_{Vl}=
g^{(0)}_{Al} -2Q_ls^2_0$).\par
Analogous expressions can be easily derived for eq.(2). We shall only
give here the relevant $Z'$ contribution, which reads:

\bq A^{(0)(Z')}_{lW}(q^2) =  {i\over
q^2-M^2_{0Z'}}({g_0\over2c_0})\bar
v_l\gamma^{\mu}(g^{'(0)}_{Vl}-\gamma_5g^{'(0)}_{Al})u_l
e_0g_{Z'WW}P_{\alpha\beta\mu}\epsilon^*_{\alpha}(p_1)\epsilon^*_{\beta}
(p_2)\eq

\noindent
where

\bq P_{\alpha\beta\mu} =
g_{\mu\beta}(2p_2+p_1)_{\alpha}+g_{\beta\alpha}(p_1-p_2)_{\mu}-
g_{\mu\alpha}(2p_1+p_2)_{\beta}    \eq

\noindent
and $p_{1,2}$ are the four-momenta of the outgoing $W^{+,-}$. In this
expression we have assumed that the $Z'WW$ vertex has the usual
Yang-Mills form. We do not consider the possibility of anomalous
magnetic or quadrupole type of couplings. In fact in most of the
popular examples based on extended gauge models \cite{Zpmod},
even with a strong
coupling regime \cite{strcoupl}, or in compositeness inspired
schemes \cite{Zpcomp}, only the Yang-Mills form appears
or at least dominates
over anomalous forms in the large $M_{Z'}/M_Z$ limit (there are however
exceptions, see for example \cite{Baur}). An
analysis with anomalous $ZWW$ and $Z'WW$ coupling forms is possible
along the lines of ref.\cite{BilWW} but is beyond the scope of this
paper.
Our analysis will be nevertheless rather general as
the trilinear $Z'WW$ coupling $g_{Z'WW}$ will be treated as a free
parameter, not necessarily proportional to the $Z-Z'$ mixing angle as
for example it would happen in a "conventional" $E_6$ picture.
\par
For the purposes of this paper, it will be particularly convenient to
describe the virtual $Z'$ effect as an "effective" modification of the
\underline{$Z$ and $\gamma$} couplings to fermions and $W$ pairs. As
one can easily derive, this corresponds to the use of the following
"modified" leptonic $\gamma$, $Z$ couplings (denoted in the following
with a star) to describe the $e^+e^-\to l^+l^-$ process:

\bq e^*_0= e_0[1-{q^2\over
M^2_{Z'}-q^2}({g^{(0)2}_{Vl}\over4s^2_0c^2_0})
(\xi_{Vl}-\xi_{Al})^2]^{1/2}
\eq

\bq g^{*(0)}_{Al} =  g^{(0)}_{Al}[1-{q^2-M^2_Z\over
M^2_{Z'}-q^2}\xi^2_{Al}]^{1/2} \eq

\bq g^{*(0)}_{Vl} =  g^{(0)}_{Vl}[1-{q^2-M^2_Z\over
M^2_{Z'}-q^2}\xi_{Al}(\xi_{Vl}-\xi_{Al})].
[1-{q^2-M^2_Z\over
M^2_{Z'}-q^2}\xi^2_{Al}]^{1/2}\eq

\noindent
and to the use of the following modified trilinear couplings
that fully describe the effect in the final $e^+e^-\to WW$ process
(without
modifying the initial $\gamma, Z$ leptonic couplings):

\bq g^{*(0)}_{\gamma WW} =  g^{(0)}_{\gamma WW}+
g^{(0)}_{Z'WW}{q^2\over M^2_{Z'}-q^2}g^{(0)}_{Vl}(\xi_{Vl}-\xi_{Al})\eq

\bq g^{*(0)}_{ZWW} =  g^{(0)}_{ZWW} -  g^{(0)}_{Z'WW}{q^2-M^2_Z\over
M^2_{Z'}-q^2}\xi_{Al}\eq

\noindent
In the previous equations, the following definitions have been used:

\bq  \xi_{Vl} =  g^{'(0)}_{Vl}/g^{(0)}_{Vl}   \eq

\bq  \xi_{Al} =  g^{'(0)}_{Al}/g^{(0)}_{Al}   \eq

\noindent
Our normalization is such that:

\bq g^{(0)}_{\gamma WW} = 1  \eq

\bq g^{(0)}_{ZWW} = c_0/s_0  \eq

\noindent
Note that, strictly speaking, only bare quantities should appear in the
previous equations. In practice, however, we shall treat the $Z'$
effect on the various observables at one loop in the ($\gamma, Z$)
Standard Model
sector, and in an "effective" tree level for what concerns the $Z'$
parameters. As a result of this (standard) approach, whose validity is
obviously related to the (implicit) assumption that $M^2_{Z'}$ is
"large" at the energy scale of the experiment, only the physical
$\gamma$, $Z$ parameters, the "physical" $Z'$ mass and the "physical"
$Z'$ couplings will remain in the various theoretical expressions. Note
that the definition of "physical" $Z'$ couplings is plagued with an
intrinsic ambiguity that would only be solved once this particle were
discovered and its decays measured. This will not represent a problem
in our approach, as we shall see, since our philosophy will rather
be that of calculating functional relationships between different
(experimentally measurable) $Z'$ shifts. In this spirit, we shall use
{}from now on only notations without "bare" indices on the various
parameters.\par
The basic idea of our approach is provided by the observation that the
modified trilinear gauge couplings eqs.(11),(12) contain the same
combinations of modified fermionic $\gamma$, $Z$ couplings eqs.(8-10)
that would appear in the leptonic final state, with only one extra free
parameter i.e. the $Z'WW$ coupling $g_{Z'WW}$. This implies that it
must be possible to find a precise relationship between the two
modified trilinear gauge couplings eqs.(11),(12) and some set of
leptonic observables by simply eliminating the free parameter
$g_{Z'WW}$ in these expressions. To fully understand what type of
relationships will emerge, it is oportune to write down at this point
the expression of the $Z'$ effects in the two leptonic
observables that will certainly be measured to a very
good accuracy at NLC, i.e. the muon ($\equiv$ lepton)
cross section $\sigma_{\mu}(q^2)$ and the muon
forward-bacward asymmetry $A_{FB,\mu}(q^2)$. To calculate these
expressions is straightforward and the rigorous derivation has been
already performed in previous papers, to which we refer for a detailed
discusssion \cite{univ}. Here we shall only show for sake of
self-completeness an approximate procedure where only the numerically
relevant terms are retained. For this aim it will be sufficient to
start from the following expressions of $\sigma_{\mu}$, $A_{FB,\mu}$ at
Born level without the $Z'$ contribution:

\bq \sigma^{(0)}_{\mu}(q^2) =  {q^2\over12\pi}[({e^2_0\over
q^2})^2+{1\over (q^2-M^2_{0Z})^2}({g^2_0\over4c^2_0})(g^{(0)2}_{Vl}+
g^{(0)2}_{Al})^2] \eq

\noindent
(we have omitted the $\gamma-Z$ interference that is numerically
negligible \cite{univ}).

\bq A^{(0)}_{FB,\mu}(q^2) =  {\pi q^2\over\sigma_{\mu}(q^2)}[
{1\over
(q^2-M^2_{0Z})^2}({1\over4\pi^2})({g^2_0\over4c^2_0})^2g^{(0)2}_{Vl}
g^{(0)2}_{Al}+{1\over q^2(q^2-M^2_{0Z})}({1\over8\pi^2})
({g^2_0\over4c^2_0})e^2_0g^{(0)2}_{Al}] \eq

\noindent
The prescription for deriving the $Z'$ effect is now the following. One
replaces the quantities $e_0$, $g^{(0)}_{Al}$,
$g^{(0)}_{Vl}$ that appear
in eqs.(17),(18) by the starred ones given in eqs.(8)-(10). All the
remaining bare ($\gamma$, $Z$) parameters will then be replaced by the
known Standard Model expressions valid at one loop,
that will contain certain
"physical" quantities and certain one-loop "corrections". For the
purposes of this paper, where only the (small) $Z'$ effects are
considered, the latter corrections can be ignored and one can write the
relevant shifts in terms of $Z'$ parameters and of $\gamma$, $Z$
"physical" quantities (more precisely, as one can guess, $\alpha(0)$
and $s^2_{Eff}(M^2_Z)$), and for a more detailed discussion we refer to
ref. \cite{univ}.\par
Defining the relative $Z'$ shifts as :

\bq {\delta\sigma^{(Z')}_{\mu}\over\sigma_{\mu}} =
{\sigma^{(\gamma,Z,Z')}_{\mu}-\sigma^{(\gamma,Z)}_{\mu}\over
\sigma^{(\gamma,Z)}_{\mu}}  \eq

\noindent
(and an analogous definition for the asymmetry),\\
it is now relatively simple
to derive the expressions:

\bq {\delta\sigma_{\mu}\over \sigma_{\mu}} = {2\over
\kappa^2(q^2-M^2_Z)^2+q^4}[\kappa^2(q^2-M^2_Z)^2
\tilde{\Delta}^{(Z')}\alpha(q^2)-q^4(R^{(Z')}(q^2)+
{1\over2}V^{(Z')}(q^2))] \eq

\bq {\delta A_{FB,\mu}\over  A_{FB,\mu}} = {q^4-\kappa^2(q^2-M^2_Z)^2
\over\kappa^2(q^2-M^2_Z)^2+q^4}[
\tilde{\Delta}^{(Z')}\alpha(q^2)+R^{(Z')}(q^2)]
+{q^4\over\kappa^2(q^2-M^2_Z)^2+q^4}V^{(Z')}(q^2)] \eq

\noindent
with (using the same notations as in
ref.\cite{univ}):

\bq  \tilde{\Delta}^{(Z')}\alpha(q^2) =
{q^2\over q^2-M^2_{Z'}}({v^2_1\over 16s^2_1c^2_1})
(\xi_{Vl}-\xi_{Al})^2
\eq

\bq R^{(Z')}(q^2) = ({q^2-M^2_{Z}
\over M^2_{Z'}-q^2})
\xi^2_{Al} \eq

\bq V^{(Z')}(q^2) =
({q^2-M^2_{Z}\over M^2_{Z'}-q^2})({v_1\over4s_1c_1})
\xi_{Al}(\xi_{Vl}-\xi_{Al}) \eq

\noindent
where

\bq   \kappa  =  {\alpha M_Z\over9\Gamma_l}\eq

\noindent
and $v_1=-2g_{Vl}=1-4s^2_1$; $s^2_1c^2_1=
{\pi\alpha\over\sqrt2G_{\mu}M^2_Z}$.\par
As one sees from the previous equations, the $Z'$ effects can be
expressed by certain quadratic expressions of the parameters
$\xi_{Al}$, ($\xi_{Vl}-\xi_{Al}$). A much simpler dependence is
exhibited by the modified trilinear couplings, as one sees from
eqs.(11),(12). Adopting the notations that are found in recent
literature \cite{GR}, \cite{BilWW},
we find for the $Z'$ effect in this case

\bq \delta^{(Z')}_{\gamma} \equiv g^*_{\gamma WW} -1=
g_{Z'WW}({q^2\over M^2_{Z'}-q^2})g_{Vl}(\xi_{Vl}-\xi_{Al}) \eq

\bq  \delta^{(Z')}_Z \equiv g^*_{ZWW} -cot\theta_W=
-g_{Z'WW}({q^2-M^2_Z\over M^2_{Z'}-q^2})\xi_{Al}  \eq

\noindent
{}From eqs.(23),(24) one can derive the following constraint:

\bq  \delta^{(Z')}_{\gamma} = tg\theta_A \delta^{(Z')}_Z   \eq

\noindent
where

\bq  tg\theta_A ={-q^2\over q^2-M^2_Z}({\xi_{Vl}-\xi_{Al}
\over \xi_{Al}})g_{Vl} \eq

A few comments are appropriate at this point. In this description both
$g_{\gamma WW}$ \underline{and} $g_{ZWW}$ couplings are modified by
form factor effects whose scale is $M_{Z'}$. They identically vanish
for $q^2=0$ and $q^2=M^2_Z$ respectively. The vanishing of
$\delta_{\gamma}$ at $q^2=0$ is absolutely required by
conservation of electric charge. We then notice that the virtual effect
of a general $Z'$ in the $WW$ channel is, at first sight, quite similar
to that of a possible model with anomalous gauge couplings, that would
also produce shifts $\delta_{\gamma}$, $\delta_Z$ \underline{both} in
the $\gamma WW$ and in the $ZWW$ couplings (in the conventional
description of anomalous gauge boson couplings the appearence of both
$\delta_{\gamma}$ and $\delta_Z$ type is
rather unusual\cite{BilWW}, but can be
described using effective lagrangians with
$dim=6$ and $dim=8$ operators \cite{GR}). For this reason,
we have called such effects "anomalous" $Z'$ effects. But the $Z'$
shifts satisfy in fact the constraint given by eq.(28), that
corresponds to a certain line in the ($\delta_{\gamma}$, $\delta_Z$)
plane whose angular coefficient is fixed by the model i.e. by the
values of $\xi_A$, ($\xi_V-\xi_A$). For example:

\bq tg\theta_A = {-q^2\over
q^2-M^2_Z}[{v_1\over2}+{cos\beta\over\sqrt{{5\over3}}sin\beta+cos\beta}]
\eq
\noindent
in $E_6$ models ($-1<cos\beta<+1$),

\bq tg\theta_A = {-q^2\over
q^2-M^2_Z}[{v_1\over2}-{2\over\alpha_{RL}}({1\over2\alpha_{RL}}
-{\alpha_{RL}\over4})]
\eq
\noindent
in Right-Left symmetric models ($\sqrt{2\over3}<\alpha_{RL}<\sqrt{2}$),

\bq tg\theta_A = {-q^2\over
q^2-M^2_Z}cot\theta_W  \eq
\noindent
in $Y$ models,

\bq tg\theta_A = {q^2\over
q^2-M^2_Z}cot\theta_W ({s^2_1-\lambda^2_Y\over1-s^2_1+\lambda^2_Y}) \eq
\noindent
in $Y_L$ models ($0<\lambda^2_Y<1-s^2_1$).\par
 The values of $\xi_A$, ($\xi_V-\xi_A$) are,
in turn, directly responsible for deviations in
the leptonic channel that would affect $\sigma_{\mu}$ and $A_{FB,\mu}$.
This means that a precise functional relationship will exist between
the value of $tg\theta_A$ defined by eq.(26) and those of the shifts
$\delta\sigma_{\mu}$, $\delta A_{FB,\mu}$ defined by eqs.(20),(21) that
would correspond to a certain surface in the 3-dim ($tg\theta_A$,
${\delta^{(Z')}\sigma_{\mu}/ \sigma_{\mu}}$,
${\delta^{(Z')}A_{FB,\mu}/ A_{FB,\mu}}$) space. To draw this surface
requires a dedicated numerical analysis that carefully takes into
account the QED initial radiation and the precise experimental set up,
which is beyond the purposes of this short paper. Here we shall only
illustrate, with a couple of particularly simple examples, what would
be typical signatures of this type of $Z'$ effects.\par
We begin with the case \cite{GA} of a $Z'$ whose couplings to the fermions are
"essentially" the same as those of the
Standard Model $Z$ (this is usually called
the "standard $Z'$ model"), leaving the $Z'WW$ coupling free. In this
case $\delta^{(Z')}_{\gamma}=0$ and $\delta^{(Z')}_Z$ is given by
eq.(24) with $\xi_{Al}$ of order one. Eqs.(19),(20) become now
\underline{at 500 GeV}:

\bq {\delta\sigma^{(Z')}_{\mu}\over \sigma_{\mu}}(\xi_{Vl}=\xi_{Al})
\simeq -0.234{q^2-M^2_Z\over M^2_{Z'}-q^2}\xi^2_{Al}   \eq

\bq  {\delta^{(Z')}A_{FB,\mu}\over A_{FB,\mu}}(\xi_{Vl}=\xi_{Al})
\simeq -0.735{q^2-M^2_Z\over M^2_{Z'}-q^2}\xi^2_{Al}   \eq

\noindent
showing that, for this situation, the asymmetry is more sensitive to
the effect. In terms of the asymmetry we have now:

\bq  \delta ^{(Z')}_Z(\xi_{Vl}=\xi_{Al}) \simeq
({g_{Z'WW}\over0.735\xi_{Al}})
{\delta^{(Z')}A_{FB,\mu}\over A_{FB,\mu}} =-g_{Z'WW}
{q^2-M^2_Z\over M^2_{Z'}-q^2}\xi_{Al}\eq

\noindent
We shall now introduce the following ansatz concerning the theoretical
expressions of $g_{Z'WW}$, that we shall write as:

\bq  g_{Z'WW} =[c{M^2_Z\over M^2_{Z'}}]cotg\theta_W  \eq

The constant $c$ would be of order one for the "conventional" models
\cite{Zpmod} where the $Z'$ couples to W only via the $Z-Z'$ mixing
(essentially
contained in the bracket). But for a general model, $c$ could be
larger, as one can see for some special cases of composite models,
for example with excited $Z^*$ states \cite{Baur} or when the $Z'$
participates in a strong coupling regime \cite{strcoupl}.\par
In fact, a stringent bound on $c$ comes from the request that the $Z'$
width into $WW$ is "small" compared to the $Z'$ mass. Inposing the
(reasonable) limit

\bq \Gamma_{Z'WW} \lsim {1\over10}M_{Z'}   \eq

\noindent
leads to the condition

\bq  c \lsim 10   \eq

\noindent
and this will be our very general working assumption.\par
Using eq.(37) we can rewrite the $Z'$ effect as

\bq   \delta^{(Z')}_Z(\xi_{Vl}=\xi_{Al}) =-\xi_{Al} {q^2-M^2_Z\over
M^2_{Z'}-q^2}c({M^2_Z\over M^2_{Z'}})cot\theta_W   \eq

\noindent
For $\xi_{Al} \simeq 1$, at the NLC energy the detectability request
\cite{BilWW}, $|\delta_Z| \gsim 10^{-2}$
corresponds to the condition:

\bq   c \gsim  2\times10^{-4}{M^2_{Z'}\over M^2_Z}[{M^2_{Z'}-(500
GeV)^2\over M^2_Z}] \eq

\noindent
which obeys the constraint eq.(32) for all values of $M_{Z'}$ such that

\bq  M_{Z'} \lsim 1.4 TeV   \eq

\noindent
and, for the limiting value $M_{Z'}=1.4 TeV$, a reasonable relative
shift of approximately ten percent in the asymmetry would be produced,
that would not be missed at the expected experimental accuracy.\par
As a second example, we consider the orthogonal case $\xi_{Al}=0$ for
"large" values of $\xi_{Vl}$ (say, $\xi_{Vl} \simeq 10$). Such an order of
magnitude corresponds to
several "conventional" models. For example in $E_6$
\bq  \xi_{Vl} = -{4s_1cos\beta\over v_1}  \eq
reaches
$\xi_{Vl} \simeq 5$ for the $\chi$ model ($cos\beta=1$),\\
and for a $Y$ model
\bq  \xi_{Vl} = {3s_1c_1\over v_1\lambda_Y}
(1-{\lambda^2_Y\over1-s^2_1})^{1/2}  \eq
reaches $\xi_{Vl} \simeq 12$ for $\lambda^2_Y=s^2_1$.
In this situation,
$\delta^{(Z')}_Z =0$, and

\bq   {\delta\sigma^{(Z')}_{\mu}\over \sigma_{\mu}}(\xi_{Al}=0)
\simeq -1.11\times 10^{-2}{q^2\over M^2_{Z'}-q^2}\xi^2_{Vl}    \eq

\bq {\delta^{(Z')}A_{FB,\mu}\over A_{FB,\mu}}  \simeq
4.71\times 10^{-3}{q^2\over M^2_{Z'}-q^2}\xi^2_{Vl}   \eq

\noindent and one sees that now the sensible quantity is
$\sigma_{\mu}$. For the trilinear shift $\delta^{(Z')}_{\gamma}$ we have
:

\bq   \delta ^{(Z')}_{\gamma}(\xi_{Al}=0) \simeq
g_{Vl}\xi_{Vl}
{q^2\over M^2_{Z'}-q^2}c({M^2_Z\over M^2_{Z'}})cot\theta_W   \eq

In correspondence to the limiting value $c=10$ and for $M_{Z'}=1.4
TeV$, we get for $\xi_{Vl}= O(10)$ a value $\delta^{(Z')}_{\gamma} =
O(10^{-2})$ i.e. the same size as $\delta^{(Z')}_Z$ in the first example
(the relative shift in $\sigma_{\mu}$ would now be of approximately
fifteen percent, that would certainly not escape experimental
detection). This value should be compared to that predicted by
dedicated analyses, that to our knowledge are still missing.\par
We have also looked whether there are reasonable possibilities to
observe such type of $Z'WW$ effects at LEP2. Assuming the lowest
allowed mass $M_{Z'}=600 GeV$ and the strongest $Z'WW$ coupling with
$c=10$ in both extreme cases studied above, one gets at most a two
percent effect in $\delta^{(Z')}_{\gamma}$ or $\delta^{(Z')}_Z$ which
is below the observability limit expected from ref.\cite{BilWW}.\par
In conclusion, we can summarize the main points of our (preliminary)
analysis as follows. A $Z'$ belonging to a quite general model, with
"reasonable" couplings to $W$ (and to fermions), would produce effects
in the $WW$ and in the leptonic channel that would be related in a
quite special way and, at least in some simple cases, visible in both
channels at a high energy $e^+e^-$ collider.
If, at the time of a possible NLC run, models with anomalous
gauge couplings will not be ruled out, this fact would certainly be a
powerful tool for discrimination. In fact, in the anomalous gauge
coupling case, the parameters that affect the $WW$ channel are totally
independent of those that affect the leptonic one so that no kind of
relationships will in general exist. If, on the contrary, anomalous
gauge couplings were out of interest, the constraints that we derived
would certainly help to achieve a proper $Z'$ identification, in case
this particle were actually produced e.g. at the CERN LHC.

\vspace{0.5cm}

{\bf \underline{Acknowledgements}}\par
This work has been partially supported by the EC contract
CHRX-CT94-0579.

\end{document}